\documentclass[aps,twocolumn,showpacs,superscriptaddress,groupedaddress]{revtex4-1}

\usepackage[english]{babel}
\usepackage[latin1]{inputenc}        
\usepackage[T1]{fontenc}
\usepackage{csquotes}

\usepackage{amsmath,amsfonts,amsthm,amsbsy}
\usepackage[svgnames,x11names]{xcolor}

\usepackage{tikz}
\usetikzlibrary{arrows}
\usetikzlibrary{intersections}

\usetikzlibrary{external}
\usepackage{pgfplots}

\usepackage{graphicx}

\definecolor{ct_red}{HTML}{E32636}
\definecolor{ct_green}{rgb}{0.47,0.67,0.19}
\definecolor{ct_blue}{rgb}{0,0.45,0.74}

\definecolor{ct2_green}{HTML}{9FF781}
\definecolor{ct2_green_dark}{HTML}{088A08}

\usepackage[pdftex,
colorlinks=true,linkcolor=blue,citecolor=blue]
{hyperref}

\newcommand{\ee}{{\mathrm e}}
\newcommand{\ii}{{\mathrm i}}
\newcommand{\dd}{{\mathrm d}}

\newcommand{\Tr}{{\mathrm{Tr}}}

\newcommand{\HB}{H_\mathrm{B}}

\newcommand{\UB}{U_\mathrm{B}}
\newcommand{\UBo}{U_\mathrm{B,1}}
\newcommand{\UBt}{U_\mathrm{B,2}}
\newcommand{\HHB}{\mathcal{H}_\mathrm{B}}

\newcommand{\HE}{H_\mathrm{E}}
\newcommand{\UE}{U_\mathrm{E}}
\newcommand{\HHE}{\mathcal{H}_\mathrm{E}}

\newcommand{\HI}{H_\mathrm{I}}
\newcommand{\UI}{U_\mathrm{I}}

\newcommand{\IB}{\mathcal{I}_\mathrm{B}}
\newcommand{\IE}{\mathcal{I}_\mathrm{E}}
\newcommand{\II}{\mathcal{I}_\mathrm{I}}

\newcommand{\vm}{\mathbf m}
\newcommand{\vn}{\mathbf n}

\begin{document}
	\title{Effective vacua for Floquet topological phases:\\ A numerical perspective on switch-function formalism}
	\date{\today}
	\author{C. Tauber}
	\affiliation{Institute for Theoretical Physics, ETH Z\"urich}
	
	\begin{abstract}
		We propose a general edge index definition for two-dimensional Floquet topological phases based on a switch-function formalism.  When the Floquet operator has a spectral gap the index covers both clean and disordered phases, anomalous or not, and does not require the bulk to be fully localized. It  is interpreted as a non-adiabatic charge pumping that is quantized when the sample is placed next to an effective vacuum. This vacuum is gap-dependent and obtained from a Floquet Hamiltonian. The choice of a vacuum provides a simple and alternative gap-selection mechanism. Inspired by the model from Rudner \textit{et al.} we then illustrate these concepts on Floquet disordered phases. Switch-function formalism is usually restricted to infinite samples in the thermodynamic limit. Here we circumvent this issue and propose a numerical implementation of the edge index that could be adapted to any bulk or edge index expressed in terms of switch functions, already existing for many topological phases.
	\end{abstract}
	
	\maketitle
	
\section{Introduction}

In the context of quantum Hall effect and topological insulators, the implementation of disorder has always played a crucial role. It ensures that a topological quantity, e.g. the Hall conductivity or a number of edge modes, remains the same for crystals that are not perfectly periodic and thus observable and invariant regardless of the microscopic impurities of a sample \cite{HasanKane10}. By analogy with  static systems, the framework of Floquet topological insulators has appeared in the last decade \cite{LindnerRefaelGalitski11,CayssolDoraSimonMoessner13}. It turns out that a periodically driven system may have topological properties when the one-period time evolution (Floquet) operator has a spectral gap \cite{KitgawaBergRudnerDemler10}. For each gap one can define a bulk topological index corresponding to a number of protected edge modes through the bulk-edge correspondence \cite{RudnerPRX13}. Moreover these indices are specific to out-of-equilibrium systems and are not entirely captured by the usual theory of static insulators, allowing for the discovery of new topological phases of matter.  

Consequently, Floquet topological phases have been studied for various symmetries and dimensions \cite{AsbothTarasinskiDelplace14,NathanRudner15,Lyon15,Lyon15bis,TauberDelplace15,Fruchart16,RoyHarper17}. The issue of disorder naturally arises also in this context, for which disordered models and their topological indices have been intensively studied recently \cite{TitumPRL15,TitumPRX16,FulgaMaksymenko16,TitumPRB17,RiederSiebererFischerFulga17}. Most of the works have been focused on the so-called Anomalous Floquet Anderson Insulator (AFAI): a new topological phase with a fully localized bulk and yet protected edge modes \cite{TitumPRX16,KunduRudnerBergLindner17,NathanRudnerLidnerBergRefael17}. But in principle disorder should be implemented for any topological phase. In this context some works from mathematical physics have generalized the bulk-edge correspondence for a large class of disordered Floquet topological insulators \cite{GrafTauber17,SadelSchulz17}. Besides disorder, recent developments have also studied the influence of interactions \cite{ElseNayak16,PotterMorimotoVishwanath16,KlinovajaStanoLoss16,HarperRoy17,RoyHarper17bis,NathanAbaninBergLindnerRudner17} and various experimental observations of these phases have been realized \cite{KitagawaPhotonicQW12,RechtsmanPhotoniFTI13,ChenPhotonicFTI14,FleuryKhanikevAlu16}.

However the physical interpretation of the topological indices in Floquet systems remains incomplete so far: In dimension two the associated observables such as charge pumping \cite{TitumPRX16} or orbital magnetization \cite{NathanRudnerLidnerBergRefael17} are  quantized only when the bulk Floquet operator is fully localized, with localization length small enough. Moreover the meaning of a spectral gap in the Floquet operator is also an open question. So far this hypothesis was made by analogy with static systems and perfectly works to define the indices, but since there is no notion of ground state in periodically driven systems it is not obvious {\it  a priori} how to select a given gap of the Floquet spectrum and observe the corresponding topological modes. Note that the AFAI phase elegantly circumvent this problem since it has one canonical gap, the bulk spectrum being completely localized. Yet the question remains open in general and is deeply related to the exciting transport properties of these systems \cite{TorresPerezBalseiroUsaj14,DehghaniOkaMitra15,FruchartDelplaceWestonWaintalCarpentier16,FarnellPereg16,EsinRudnerRefaelLindner17,KolodrubetzNathanGazitMorimotoMoore17}.

\begin{figure}[htb]
	\centering
	\includegraphics{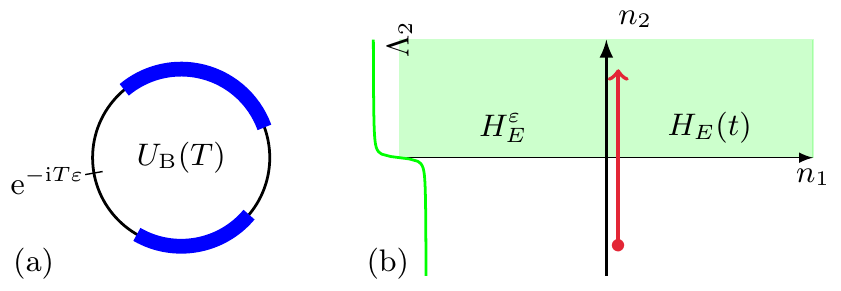}
	\caption{\label{fig:intro} (a) Each $\varepsilon$ inside a gap of the bulk Floquet  spectrum  corresponds to an effective Hamiltonian $\HE^\varepsilon$ playing the role of an effective vacuum. (b) When the original sample is placed next to this vacuum, a  quantized pumping of charge occurs at their interface within a driving cycle. The green curve $\Lambda_2$ is a switch function from $0$ to $1$ in $n_2$-direction measuring the density of electrons in the upper half-space. It plays a central role in the construction of the indices. }
\end{figure}

This work focuses on the study of the edge index for a general Floquet topological system in dimension two. We show that the interpretation of a quantized pumping within a cycle, already studied in the AFAI phase \cite{TitumPRX16}, is actually valid for any Floquet phase with a spectral gap (e.g. Fig~\ref{fig:intro}(a)). This covers both clean and disordered phases, anomalous or not, without any assumption on localization. This pumping is observed relatively to an effective dynamics given by the so-called effective Hamiltonian \cite{KitgawaBergRudnerDemler10}, that depends on the gap of the Floquet operator. Alternatively, we show that this phenomenon can be also observed when placing the effective dynamics next to the original one, namely at the interface between the two (Fig.\,\ref{fig:intro}(b)). In this setting the effective Hamiltonian appears as an effective vacuum that compensates other contributions from the bulk. There is actually one effective vacuum per gap so that each vacuum provides a way to select a gap and the corresponding edge modes. We illustrate these statements on a two-band model generated by a piecewise constant Hamiltonian introduced by Rudner \textit{et al.} in \cite{RudnerPRX13} and add a disordered on-site potential similarly to \cite{TitumPRX16} but in different regimes : first in the anomalous phase at weak disorder, where localization lentgh is large and then in a non-anomalous phase.

These results follow from a general proof of bulk-edge correspondence for Floquet Topological insulators, where the consequences on the edge index where not investigated in details \cite{GrafTauber17}. It is based on functional analysis techniques that were first developed in the context of the Quantum Hall Effect \cite{AvronSeilerSimon94}. The central notion in this framework is the switch function, that is very elementary and allows to define indices without requiring disorder averaging, large-time limit or ergodicity. In particular the introduction of external fluxes threaded through the sample is not needed and makes our approach simpler.

The formalism of switch functions is a powerful tool of mathematical physics that goes way beyond periodically driven systems and naturally appears in the context of linear response theory for topological phases \cite{AvronSeilerSimon94,ElgartSchlein04,EGS05,BoucletGerminetKleinSchenker05}. It is also used here and there in the physics literature \cite{Kitaev08,FidkowskiJacksonKlich11} (probably more often, without naming it) but somehow underestimated to compute topological indices. Its main inconvenient is that it works only for infinite samples in the thermodynamic limit. Thus it might appear cumbersome to manipulate. We propose a solution to this issue by providing a numerical implementation of this formalism on finite-size systems. We define an approximate index that coincides with the exact one in the infinite size limit. Although other rigorous indices already exist for disordered models, including a numerical implementation \cite{Prodan11, Prodan17}, our approach does not require much knowledge of the mathematical machinery behind. This makes the switch function very appealing in order to generalize and compute topological indices for disordered models. Although we illustrate this numerical implementation for the aforementioned edge index, it can be adapted to any bulk or edge index in principle.

The paper is organized as follows: in Sect.\,\ref{sec:quantized_pumping} we define the edge index and study its physical interpretation. In Sect.\,\ref{sec:interface} we present a dual picture in terms of an interface index and introduce the effective vacua and their physical properties. Then in Sect.\,\ref{sec:num} we implement these notions in a numerical framework with an application to a specific model. Sect.~\ref{sec:ccl} concludes and discusses several interesting perspectives of this work.

\section{Edge index and quantized pumping \label{sec:quantized_pumping}}

\subsection{Floquet topological insulators}

\paragraph{The bulk picture} The concepts from this section involve simple mathematical expressions but the price to pay is to deal with infinite or semi-infinite spaces. Thus we define the bulk Hilbert space $\HHB= \ell^2(\mathbb Z^2)$ for which a state $\psi$ is defined by the amplitude $\psi_{\vm}$ on each site $\vm=(m_1,m_2) \in \mathbb Z^2$ of the lattice. Internal degrees of freedom (spin, sub-lattice,...) can be taken into account by considering $\ell^2(\mathbb Z^2) \otimes \mathbb C^N$ instead, but for simplicity we focus on $\HHB$ (\textit{i.e.} $N=1$) below. An operator $A_\mathrm{B}$ on $\HHB$ can be thought as an infinite matrix $(A_\mathrm{B})_{\vm,\vn}$ for $\vm,\vn \in \mathbb Z \times \mathbb Z$.  

The initial input is a bulk time-periodic Hamiltonian $\HB(t+T)=\HB(t)$, namely a family of infinite matrices $\HB(t)_{\vm,\vn}$ for $t \in [0,T]$. The only requirement for the following formalism to work is that $\HB$ is \textit{local}:
\begin{equation}\label{local}
|\HB(t)_{\vm,\vn} | \leq C \ee^{-\mu|\vm-\vn|},
\end{equation}
for some $C,\,\mu >$ and independent of $t$. This property is also called \textit{short-range}, with range $1/\mu$, or \textit{near-sighted} \cite{ProdanKohn205}. It means that the dynamics of a state on some site is mostly ruled by its amplitude within a small neighborhood of it. Note that this does \emph{not} imply that the system is in a localized regime.  The simplest local example is a translation-invariant system where $(\HB)_{\vm,\vn} = (\HB)_{0,\vm-\vn}$: For a sample with finite range hopping $r$ (e.g. 1 for nearest neighbor), the latter vanishes for $|\vm-\vn|>r$ so that $\eqref{local}$ is trivially satisfied. However requiring \eqref{local} allows to consider any disordered configuration (on-site potential, disordered hopping, ...). 

\begin{figure}[htb]
	\includegraphics{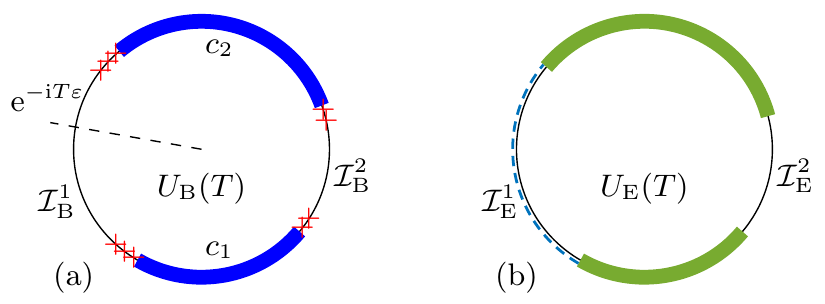}
	\caption{(a) A bulk spectrum with two bands, containing delocalized waves (blue)  and localized states (red crosses), and two spectral gaps. For each gap one defines a bulk invariant $\IB^i$ that is related to the others by the Chern number of the bands. (b) Corresponding edge spectrum: two analogous bands are present but each gap might be filled by some modes confined at the edge of the sample and counted by $\IE^i$. In this example $\IB^1=\IE^1=1$, $\IB^2=\IE^2=0$ and $c_1=-c_2=1$. \label{fig:bulk_edge_spectra}}
\end{figure}

Time-periodic dynamics can be solved through the Floquet formalism by considering the spectral properties of time evolution operator $\UB(t)$ (or unitary propagator) computed by solving the Schrödinger equation $\ii \partial_t \UB(t) = \HB(t) \UB(t)$ and $\UB(0) = 1$, or alternatively by using a time-ordered exponential of the integral of $\HB(t)$ over time. It was noticed in \cite{KitgawaBergRudnerDemler10,RudnerPRX13} that the crucial assumption to define a topological quantity is that the one-period propagator $\UB(T)$ has a spectral gap. Since it is unitary, the spectrum of $\UB(T)$ lies in the complex circle and its eigenvalues $\varepsilon$ for which $\UB(T) \psi = \ee^{-\ii T \varepsilon} \psi$ are called quasi-energies since they are defined modulo $2\pi/T$, by analogy with Bloch quasi-momenta. The quasi-energies of $\UB(T)$ are usually plotted in terms of Bloch bands over the Brillouin torus for a translation-invariant system. But we could instead project all these quasi-energy bands onto the unit circle while varying quasi-momentum. The latter description of the spectrum of $\UB(T)$ is still available when translation invariance is broken. A typical situation illustrated in Fig.\,\ref{fig:bulk_edge_spectra}(a). It is composed of several bands that contain both delocalized waves (corresponding to Bloch waves in a clean system) and localized states due to disorder. We assume that there exists at least one gap in the spectrum.

For each quasi-energy $\varepsilon$ corresponding to a spectral gap of $\UB(T)$ it is possible to define a topological invariant by using the effective Hamiltonian
\begin{equation}\label{defHeff}
\HB^\varepsilon = \dfrac{\ii}{T} \ln_{-\varepsilon T} \big( \UB(T)\big)
\end{equation} 
namely the logarithm of $\UB(T)$ with a branch cut taken in the chosen gap. This effective Hamiltonian was introduced to construct a relative evolution that produces a time-periodic evolution and allows for the definition of a topological bulk index $\IB(\varepsilon)$ \cite{RudnerPRX13,GrafTauber17}. This index is independent of $\varepsilon$ inside a given gap. Furthermore $\IB(\varepsilon')-\IB(\varepsilon)=c(P)$ where $c$ is the Chern number of the projection to the band of $\UB(T)$ between $\ee^{-\ii T \varepsilon}$ and $\ee^{-\ii T \varepsilon'}$ clockwise. 

In this paper instead we would like to discuss the physical interpretation of effective Hamiltonian \eqref{defHeff}. First it is time-independent so its dynamics is simpler than the original driven system $\HB(t)$. Second, it was proved in \cite{GrafTauber17} that $\HB^\varepsilon$ satisfies \eqref{local}, namely it is also local (but \emph{not} necessarily localized). This is because for a finite duration $t \in [0,T]$, time evolution $\UB(t)$ is also local. Finally the effective Hamiltonian has a natural interpretation in the edge picture that we  investigate now.

\paragraph{The edge picture} The edge Hilbert space is $\HHE= \ell^2(\mathbb N \times \mathbb Z)$ and describes the right half-space, namely a lattice with $m_1 \geq 0$ and a single vertical edge at $m_1 =0$. States and operators on $\HHE$ are similarly described than in $\HHB$ except that one direction is semi-infinite only. Any state $\varphi \in \HHE$ can be embedded into $\HHB$ by setting to zero the components on the left half-space and any $\psi \in \HHB$ can be truncated to a state in $\HHE$ by forgetting the components outside. For an operator $A_\mathrm{B}$ we denote by $\widehat{A_\mathrm{B}}$ the corresponding truncated operator on $\HHE$. This might be thought as taking a upper-left block of the matrix $(A_\mathrm{B})_{\vm,\vn}$, keeping only $m_1$ and $n_1 \geq 0$, and corresponds physically to the Dirichlet boundary condition. The edge Hamiltonian is then defined as the truncation of the bulk one, i.e. $\HE(t) := \widehat{\HB}(t)$. By construction it is also $T$-periodic and local. Moreover it generates a time evolution $\UE(t)$ that is unitary on $\HHE$. However note that $\UE(t) \neq \widehat{\UB}(t)$. These two operators cannot be equal since $\widehat{\UB}$ is not unitary anymore: the truncation procedure forgets some information. In other words the operations of truncation and generating time evolution do not commute, so that $\HB \mapsto \widehat{\HB} =\HE \mapsto \UE$ is not the same as $\HB \mapsto \UB \mapsto \widehat{\UB}$.

The spectrum of $\UE(T)$ is illustrated in Fig.\,\ref{fig:bulk_edge_spectra}(b). The original gaps of $\UB(T)$ may be filled with modes that are confined at the edge of the sample. The presence of these modes is characterized by an edge index $\IE(\varepsilon)$. Although $\IE(\varepsilon)=\IB(\varepsilon)$ by the bulk-edge correspondence \cite{GrafTauber17}, it is actually interesting to look at its own expression that has a nice interpretation in Floquet topological phases.

\subsection{Periodic time evolution}

Let us assume first that time-evolution is periodic, namely $\UB(T)=1$. There is a canonical spectral gap, namely every point of the circle except 1. In that case the edge invariant is defined by
\begin{equation}\label{defIE}
\IE=\Tr\Big(\UE^*(T) \Lambda_2 \UE(T)- \Lambda_2\Big) \in \mathbb Z
\end{equation}
where the trace is performed over $\HHE$ and $\Lambda_2$ is called a switch function operator \cite{AvronSeilerSimon94}. It is a diagonal operator that is defined by $(\Lambda_2)_{\vm,\vn} = \delta_{\vm,\vn} f(n_2)$ with
\begin{equation}\label{defswitchf}
	f(n_2) = \left\lbrace\begin{array}{lll}
	1 & \mathrm{for} & n_2 \geq 0\\
	0& \mathrm{for} & n_2 < 0
	\end{array}\right.
\end{equation}
Operator $\Lambda_2$ is the (infinite) density of electrons in the upper-right quadrant of the edge space. In principle we can take any switch as long as $f$ is 1 (resp. 0) for $n_2$ large and positive (resp. negative), but we stay with the previous example for concreteness. The first thing to notice is that expression \eqref{defIE} of $\IE$ is not trivially vanishing. Because $\HHE$ is infinite dimensional there is no reason that when splitting the trace each part remains finite (indeed the trace of $\Lambda_2$ is $+\infty$), so we cannot use cyclicity separately. The mathematical properties of $\IE$ were studied in \cite{GrafTauber17}, but the fact that $\IE \in \mathbb Z$ can actually be checked numerically, see Sect.~\ref{sec:IEnum} below. 


The physical interpretation of $\IE$ is the following: In the Heisenberg picture $\UE^*(T) \Lambda_2 \UE(T)- \Lambda_2$ is the relative density of electrons in the upper quadrant of the edge space between $t=0$ (where $\UE(0)=1$) and $t=T$. Even if each density is separately infinite, the difference is finite and actually quantized. Thus $\IE$ counts the (algebraic) number of electrons that have been pumped from the lower to the upper quadrant within a cycle (see also \cite[Fig. 2]{GrafTauber17}). This pumping is quantized and actually confined along the edge. Indeed, one has
\begin{equation}\label{UEvstUB}
\UE(T) = \widehat{\UB}(T) + D
\end{equation}
with $|D_{\vm,\vn}|\leq \mathcal D \ee^{-\alpha |n_1|}$ for $\mathcal D, \, \alpha >0$, so that $D \simeq 0$ as soon as $n_1$ is large. This corresponds to the fact that for $t \in [0,T]$, $\UE \simeq \widehat{\UB}$ away from the edge. When $\widehat{\UB}(T)=1$ then expression \eqref{defIE} only involves $D$ that is confined near the edge.

When the system is translation invariant, one applies Bloch decomposition in direction 2 and show that the edge index becomes \cite{GrafTauber17}
\begin{equation}\label{IEwinding}
\mathcal I_E = \dfrac{1}{2\pi \ii } \int_0^{2\pi} \dd k_2 \Tr(\UE^*(T,k_2) \partial_{k_2} \UE(T,k_2))
\end{equation}
where the trace is performed along the remaining semi-infinite direction 1. This formula is analogous to the one in \cite{RudnerPRX13}. When $\UB(T)=1$ then its spectrum is fully degenerated to a single point $\{1 \}$ so that $\IE$ can be seen as the winding number of $\UE(T)$ of eventual edges states appearing around the circle in its spectrum. When Bloch momentum $k_2$ is not available, definition \eqref{defIE} of $\IE$ then appears as a generalized (or non-commutative) winding number that can be computed even for disordered configurations.

\subsection{General case}

When $\UB(T) =1$ there is only one canonical gap that is the circle with $\{1\}$ excluded. In the general case where $\UB(T) \neq 1$  definition \eqref{defIE} is not valid anymore, but one can define an edge invariant for each spectral gap of $\UB(T)$. The problem when $\UB(T)\neq 1$ is that the operator involved in \eqref{defIE} is not confined near the edge so that its trace is not finite. Nevertheless \eqref{UEvstUB} is still true, except that now $\widehat{\UB}(T)$ might contain states infinitely far from the edge that also contribute to the pumping, leading to an infinite quantity. Thus this contribution should be somehow subtracted in order to recover a proper pumping confined at the edge.

The bulk invariant was originally defined by constructing a relative evolution generated by a dynamics with $\HB(t)$ for the first half of the period and using the effective Hamiltonian  \eqref{defHeff} for the second half \cite{RudnerPRX13,GrafTauber17}. Here we propose the same procedure but in the edge picture.
\begin{equation}\label{defHErel}
\HE^\mathrm{rel}(t) := \left\lbrace\begin{array}{lll}
2\HE(2t) & \mathrm{for} & 0 \leq t \leq T/2\\
- 2 \HE^\varepsilon & \mathrm{for} & T/2 \leq t \leq T
\end{array}\right. 
\end{equation}
where $\HE^\varepsilon := \widehat{\HB^\varepsilon}$. Note that the bulk analogue $\HB^\mathrm{rel}$ (before truncation) generates a time evolution that satisfies $\UB^\mathrm{rel}(T)=1$ and was originally constructed for this purpose. Thus we are back to the previous case and can apply definition \eqref{defIE} for $\UE^\mathrm{rel}(T)$ instead of $\UE(T)$. One important point is that $\HE^\varepsilon$ is not the logarithm of $\UE(T)$ (that might even be not gapped) but only the truncation of $\HB^\varepsilon$ defined in \eqref{defHeff}. Up to a small computation postponed to App.~\ref{app:IEeps}, we infer the edge index expression
\begin{equation}\label{defIEeps}
\IE(\varepsilon)=\Tr\Big(\UE^*(T)\Lambda_2 \UE(T)- \ee^{\ii T \HE^\varepsilon}\Lambda_2\ee^{-\ii T \HE^\varepsilon} \Big) \in \mathbb Z.
\end{equation}
As before, we get a quantized pumping within a cycle, but relatively to the dynamics due to $\HE^\varepsilon$. Note that by construction $\UB(T)=\ee^{-\ii T \HB^\varepsilon}$ so  both $\UE(T)$ and $\ee^{-\ii T \HE^\varepsilon}$ satisfy \eqref{UEvstUB} with different $D$ but with the same $\UB(T)$, namely they coincide away from the edge. By itself the pumping associated to $\UE(T)$ is not well defined because of an infinite contribution from the bulk. This contribution is removed by the pumping due to $\HE^\varepsilon$ so that the relative pumping is well defined. Note that the spectrum of $\HE^\varepsilon$ might include (truncated) delocalized waves, localized states and even edge modes from other gaps, see Sect.\,\ref{sec:eff_vac} below.

\section{Interface picture \label{sec:interface}}

When $\UB(T) \neq 1$ the invariant is defined by \eqref{defIEeps} through the regularization by $\HE^\varepsilon$ that relies on the choice of a gap $\varepsilon$. Although mathematically well defined, the physical interpretation of $\IE$ remains unpleasant because the implementation of the relative dynamics \eqref{defHErel} in an experiment might be laborious as one has to switch alternatively the physical and effective dynamics. Here we would like to propose a dual picture that provides a simpler interpretation of $\HE^\varepsilon$. Instead of a dynamics relative in time consider one relative in space, namely an interface defined by
\begin{equation}\label{sharpH}
H_\#(t)_{\vm,\vn} := \left\lbrace\begin{array}{lll}
\HE(t)_{\vm,\vn} & \mathrm{for} & m_1,n_1 \geq 0\\
(\HE^\varepsilon)_{\vm,\vn} & \mathrm{for} & m_1,n_1 < 0
\end{array}\right. 
\end{equation}
compare with \eqref{defHErel}. This Hamiltonian is a gluing inside the bulk space of $\HE(t)$ on the right half-space ($n_1 \geq 0)$ and $\HE^\varepsilon$ on the left one ($n_1 <0$). So far this is only a sharp interface where the two halves are not connected (in other words, the matrix of $H_\#$ is block diagonal in the basis $\{n_1 \geq 0, n_1 <0\}$). Thus we consider a more general interface Hamiltonian $\HI(t) := H_\#(t) + H_\mathrm{int}(t)$ where the latter part allows for a smoother gluing at the interface. We require that this perturbation stays confined at the interface, namely $|H_\mathrm{int}(t)_{\vm,\vn}| \leq A \ee^{-\alpha |n_1|}$ for some $A, \alpha >0$. This implies that $H_\mathrm{int}(t)$ vanishes quickly away from the interface $n_1=0$. Note that this condition is similar to the property of $D$ given below \eqref{UEvstUB} except that $H_\mathrm{int}(t)$ acts on $\HHB$ instead of $\HHE$ so that it is confined at both sides of the interface. Moreover the notion of confinement is much stronger than the notion of locality defined in \eqref{local}. The latter implies that the off-diagonal elements decay exponentially when the distance $|\vm-\vn|$ grows, whereas for a confined operator all the matrix elements decay exponentially as soon as one coordinate is far from the interface or the edge.

The interface Hamiltonian also generates a time evolution $\UI(t)$ on $\HHB$. This evolution has no reason to be $1$ at $t=T$, however we know the dynamics far away from the interface, namely $\UI(T) \simeq \UE(T)$ far away to the right and and $\UI(T) \simeq \ee^{-\ii T \HE^\varepsilon}$ far away to the left. Moreover we know that $\UE(T) \simeq \UB(T)$ and $\ee^{-\ii T \HE^\varepsilon} \simeq \UB(T)$ far away from the edge, so that finally $\UI(T) \simeq \UB(T)$ far away from both sides of the interface. We define the associated interface index by
\begin{equation}\label{defII}
\II(\varepsilon) = \Tr\big(  \UI^*(T)\Lambda_2\UI(T) - \UB^*(T)\Lambda_2 \UB(T)  \big) \in \mathbb Z
\end{equation}
Note that here the trace is performed over the bulk space $\HHB$. The normalization by $\UB(T)$ is of similar kind of the one discussed above by the introduction of $\ee^{-\ii T \HE^\varepsilon}$: it removes the undesirable contribution. Since $\UI(T)$ and $\UB(T)$ coincide far away from the interface and by definition nothing happens near the interface for the bulk evolution $\UB$, then $\II$ measures the quantized pumping confined at the interface between physical evolution $\HE$ and the effective one $\HE^\varepsilon$.


\subsection{Effective Vacua \label{sec:eff_vac}} The interface picture provides a dual picture where the quantized pumping arises between a physical and an effective sample placed next to each other (see \cite[Fig. 5]{GrafTauber17}). The topological nature of the index $\II$ ensures that it remains the same regardless of the gluing condition $H_\mathrm{int}(t)$ at the interface. Moreover we argue in App.~\ref{app:IE=II}  that 
\begin{equation}\label{IE=II}
 \IE(\varepsilon) = \II(\varepsilon)
\end{equation}
so that the two pictures are equivalent. Remember that $\HE^\varepsilon = \widehat {\HB^\varepsilon}$ is the edge truncation of the bulk effective Hamiltonian. In the particular case where $\UB(T) = 1$ there is a canonical spectral gap and $\HB^\varepsilon $ vanishes for every $\varepsilon \in (0,2\pi)$, so that $\HE^\varepsilon=0$ and \eqref{defIE} coincides with \eqref{defIEeps}. In that case the pumping is quantized relatively to the vacuum $\HE^\varepsilon = 0$. In the general case, $\HE^\varepsilon$ plays the role of an effective vacuum that allows for a quantized pumping, equivalently in a relative dynamics or at the interface with it. 

Effective vacua $\HE^\varepsilon$ have the following interesting properties, mostly inherited from $\HB^\varepsilon$. They are: (i) time-independent, (ii) local in the sense of \eqref{local} and (iii) independent of $\varepsilon$ inside a given gap of $\UB(T)$.  The latter property tells that there are as many distinct effective vacua as spectral gaps in $\UB(T)$. Because each vacuum comes from a logarithm of $\UB(T)$, it might be difficult to compute it explicitly in general. Nevertheless, because of property (ii) it can always be well approximated by a finite range Hamiltonian. Furthermore it might be computed numerically, see below. Finally, as we shall see its spectrum is known exactly and characterized by the one of $\UB(T)$, and each vacuum allows to select one specific gap, by analogy with a choice of chemical potential through a particle reservoir in static topological insulators. 

\begin{figure}[htb]
	\includegraphics{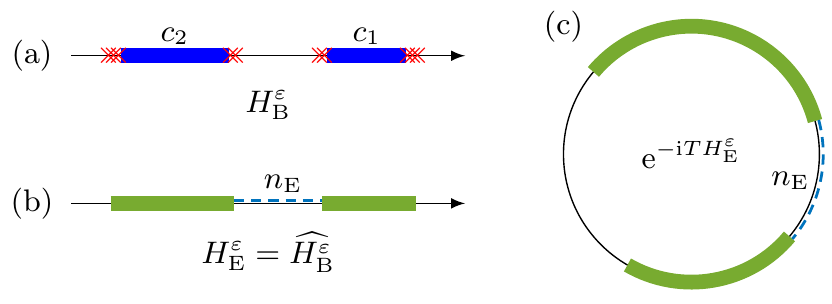}
	\caption{(a) The spectrum of $\HB^\varepsilon$ is obtained by unwinding the one of $\UB(T)$ on the real line, clockwise. The gap of the branch cut is split into two parts below and above the real spectrum, and the other bands and gaps are preserved. (b) The truncation $\HE^\varepsilon$ on the edge space may have edge modes inside the remaining gap accordingly to the Chern number of $\HB^\varepsilon$ below this gap. (c) By construction $\ee^{-\ii T \HE^\varepsilon}$ is always gapped around the original branch cut and may have edge modes in the other gaps according to the value of the other invariants. Following the example of Fig.\,\ref{fig:bulk_edge_spectra}, we get $n_\mathrm{E} = c_2= -1$ here. \label{fig:effective_spectrum}}
\end{figure}

Imagine a standard situation with two bands and two gaps  (Fig.\,\ref{fig:bulk_edge_spectra}(a)), that is straightforward to generalize. The gap of $\UB(T)$ chosen around $\varepsilon$ is split into two parts and the rest of the spectrum is unwinded from the circle to the real line, giving the spectrum of $\HB^\varepsilon$, see Fig.\,\ref{fig:effective_spectrum}(a). After truncation, $\HE^\varepsilon$ may have edge modes in a remaining gap, its number being equal to the chern number of the band below the gap, by the usual bulk-edge correspondence of static topological insulators, see Fig.\,\ref{fig:effective_spectrum}(b). Note that these Chern numbers come from the bands of $\UB(T)$. Then by folding again the spectrum of $\HE^\varepsilon$ around the circle we get the one for $\ee^{-\ii T\HE^\varepsilon}$ (Fig. \ref{fig:effective_spectrum}(c)) that has to be compared with $\UE(T)$ (Fig. \ref{fig:bulk_edge_spectra}(b)) : they have the same bands but not the same edge modes. In particular $\ee^{-\ii T\HE^\varepsilon}$ has no edge mode in the gap around $\varepsilon$ by construction.

Note that in the particular case of so-called anomalous phases where the edge indices are the same in each gap and all the Chern numbers are vanishing \cite{RudnerPRX13}, then the effective vacuum has no edge mode at all so it is topologically trivial. In a more general case, it contains some topological edge modes that are also required in the regularization process, in order to have a quantized pumping in expression \eqref{defIEeps} for $\IE$ or \eqref{defII} for $\II$. When $\UB(T)$ has several gaps with distinct indices, part of the topology has to be \enquote{removed} in order to select one gap associated to one specific value and observe the corresponding quantized pumping. This is the role of $\HE^\varepsilon$ so it is not surprising that it is topological in general. However for the anomalous phases the invariant has the same value in each gap, and only the contribution from delocalized states is removed through $\HE^\varepsilon$, hence its trivial topology.

\section{Numerical implementation \label{sec:num}}

In Sect.\,\ref{sec:quantized_pumping} we have defined an edge index valid for any disordered configuration of the sample without any average over disorder, that allows for an efficient way to compute the index and characterizes the topology of the system. The major inconvenient of this framework is to deal with infinite systems. If everything is mathematically correct, it seems rather difficult to implement in an experiment or even numerically. For example, expression \eqref{defIE} trivially vanishes if the Hilbert space is finite dimensional by cyclicity of the trace. Here we show that these problems can be circumvented and that there is a way to estimate numerically the previous indices. We illustrate by the way the different statements from Sect.\,\ref{sec:quantized_pumping} and \ref{sec:interface}. The code to generate all the figures below is available as a supplementary material of this paper.

\subsection{The model \label{sec:model}}  

We use a two-band model first proposed by Rudner \textit{et al.} with translation invariance \cite{RudnerPRX13}, for which an experimental realization in optical lattices was recently proposed \cite{QuelleWeitenbergSengstockSmith17}. Then disorder was included in \cite{TitumPRX16} to generate the anomalous Floquet-Anderson insulator phase where all the states of the bulk Floquet operator are localized. Here we also include disorder but rather consider a generic situation with delocalized bands and localized states. The crystal is a bipartite square lattice divided into two sublattices $A$ and $B$. The Hamiltonian is time-periodic with period $T$ and piecewise constant in time: $H(t) = H_n$ for $(n-1)T/5 \leq t < n T/5$ with $n\in \{1,\ldots,5\}$. The  first four steps are hopping terms with a common hopping parameter $J$ where the different bonds of the bipartite lattice are alternatively switched on and off. Each step connects $A$-sites with nearest neighbor $B$-sites, respectively situated to the right, top, left, and bottom. The last step is a pure on-site potential (no hopping) with a disordered potential
\begin{equation}\label{defH5}
(H_5)_{\vm,\vn} = \pm (\delta+\delta_rV_\vm) \delta_{\vm,\vn} 
\end{equation}
where $\pm$ refers to $A$ or $B$ sites, $\delta, \delta_r \in \mathbb R$ and $\{V_\vm\}$ is a uniform (identically distributed) random variable with support in $[-1/2, 1/2]$. In following we choose $\delta=\delta_r$ for simplicity. Other probability distributions can be also implemented. 
 
\begin{figure}[htb]
	\includegraphics{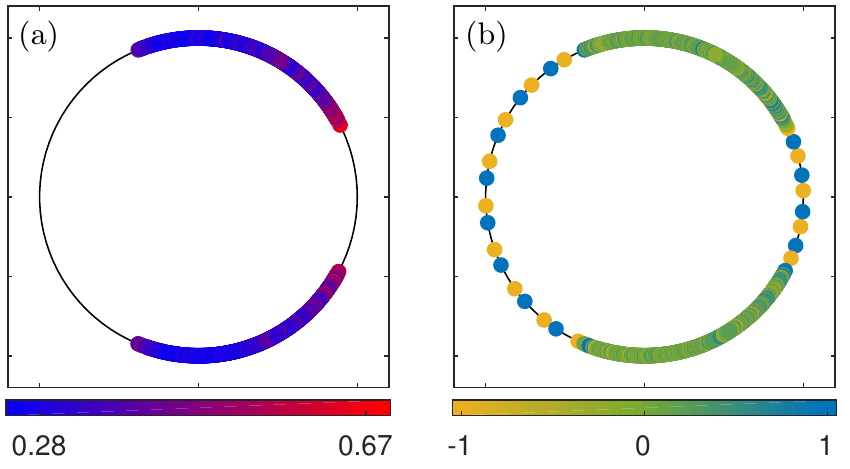}
	\caption{\label{fig:num_spectra} Numerical spectra of a disordered configuration for $J=1.875\pi/T$, $\delta=1.6\pi/T$ and $L=40$. (a) Bulk spectrum with two bands and two gaps. The colorbar indicates the participation ratio $\alpha$ of the corresponding eigenstate. As expected each band is globally delocalized ($\alpha \sim 0.3$) with localized states at the extremities ($\alpha \sim 0.7$). (b) Edge spectrum: the gaps are filled with edge states. The colorbar indicates the first momentum $\beta$ in direction $1$ of the corresponding eigenstate's probability density. Edges states are confined to the left (resp. right) of the sample when $\beta=-1$ (resp. $\beta=1$), whereas the bands are not particularly confined ($\beta \sim 0$).}
\end{figure}

We now work on a finite size square sample $\vn \in [1,L]^2$ and distinguish the $A$ and $B$-sites by the parity of the coordinates $(n_1,n_2)$. Whatever the boundary conditions are, the time-evolution after one period is given by
\begin{equation}
U(T) = \ee^{-\ii \frac{T}{5} H_5} \cdots \ee^{-\ii \frac{T}{5} H_1}
\end{equation}
Depending on the value  of $J$ and $\delta$, the model presents various configurations of bands and edge modes as we shall see. The bulk Hamiltonian is given by $\HB=H$ with periodic boundary conditions in both directions, and the corresponding bulk evolution $\UB(T)$ has a typical spectrum of two bands and two gaps illustrated in Fig.\,\ref{fig:num_spectra}(a). For each eigenvalue $\lambda$ we compute the inverse participation ratio of the (normalized) eigenstate $\psi^\lambda$
\begin{equation}
\alpha = \sum_{\vn \in [1,L]^2} |\psi^\lambda_\vn|^4
\end{equation}
If the state is perfectly localized at a given site $\vn_0$, $\psi^\lambda_\vn = \delta_{\vn,\vn_0}$, then $\alpha =1$ and if the state is completely delocalized then $\psi^\lambda_\vm = 1/L$ and $\alpha = 1/L^2 \rightarrow 0$ in the thermodynamic limit. As expected we observe two bands that are mostly delocalized, with some localized states at their extremities. Compare with Fig.\,\ref{fig:bulk_edge_spectra}(a).
 
The edge Hamiltonian is given by $\HE = H$ with Dirichlet boundary condition in direction 1 and periodic boundary condition in direction 2, so that the crystal is a cylinder with two edges at $\vn_1 = 1 $ and $\vn_1=L$.  The corresponding edge evolution $\UE(T)$ has a typical spectrum illustrated in Fig.\,\ref{fig:num_spectra}(b). Similarly to the bulk spectrum one has two bands but the gaps are now filled by modes that are confined at one of the two edges. For each eigenvalue $\lambda$ we compute the first momentum in direction 1 of the corresponding eigenstate's probability density
\begin{equation}
\beta = \sum_{\vn \in [1,L]^2} \frac{2n_1-L-1}{L-1}|\psi^\lambda_\vn|^2
\end{equation}
If $\psi^\lambda$ is fully confined at $n_1=1$ (resp. $n_1=L$) then $\beta = -1$ (resp. $\beta =1$). If it is completely delocalized then $\beta =0$ by parity around $L/2$. As expected we observe edge modes between delocalized bands. The difference with Fig.\,\ref{fig:bulk_edge_spectra}(b) is first that we have two edges so two locations for the edge modes and then that both gaps are filled with edge modes. We are indeed in an anomalous phase.

For a fixed $\delta$ the phase of the system is $5\pi/T$-periodic in $J$, even with respect to $J=2.5\pi/T$ so we restrict our analysis to $J \in [0,2.5\pi/T]$. At $J=0$ one has $\UB(T) = 1$ and the system is topologically trivial until $J^* = 1.25 \pi$ where both gaps close, leading to the anomalous phase with one edge mode in each gap. Finally at $J=2.5\pi/T$ one also has $\UB(T)=1$.

\subsection{Switch functions and periodic boundary condition \label{sec:switch_per}}

The first problem encountered in the numerical computation of the edge index is that we necessarily have two edges, leading to two counter-propagating edge modes and a vanishing index. This explains why \eqref{defIE} has to vanish on a finite size sample. This situation is quite common in the computation of topological invariants, and the usual solution is to introduce a cut-off. By performing the trace over the left half part of the sample only, namely $n_1 \in [1,L/2]$, we expect to estimate the index associated to one edge only, that is non vanishing and coincides with the previous definition of $\IE$ in the thermodynamic limit.

\begin{figure}[htb]
	\centering
	\includegraphics{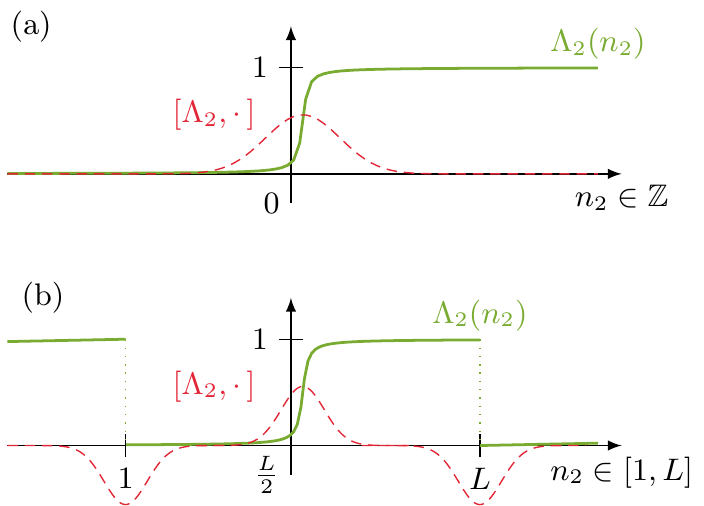}	
	\caption{(a) For an infinite sample a switch function is any function that switches between 0 and 1 from $-\infty$ to $+\infty$. Operator $[\Lambda_2, \cdot]$ is confined near in the region of the switch, required to construct finite-trace expressions. (b) In a finite sample with periodic boundary condition, $[\Lambda_2, \cdot]$ has two opposite contributions because of periodicity, leading to a vanishing total invariant. \label{fig:switch_functions}}
\end{figure}

We actually have the same issue here in direction 2 because of the switch function $\Lambda_2$. In the infinite setting, a switch function is given by any function that is 1 (resp. 0) for $n_2$ positive (resp. negative) and large. Equation \eqref{defswitchf} is just one example, but the index $\IE$ is actually independent of the choice of switch function \cite{GrafTauber17}. Noticing that \eqref{defIE} can be rewritten
\begin{equation}
\IE = \Tr(\UE^*(T)[\Lambda_2,\UE(T)])
\end{equation}
it is actually possible to show that $[\Lambda_2, \cdot]$ acts as some kind of non-commutative derivative (compare with \eqref{IEwinding}), so that intuitively $[\Lambda_2, \UE(T)]$ is significant only where $\Lambda_2$ varies: near the switching $n_2=0$ (Fig.\,\ref{fig:switch_functions}(a)). This actually ensures that the trace of $\UE(T)[\Lambda_2, \UE(T)]$ is finite and $\IE$ is well-defined in the infinite setting. However, when we work with periodic boundary conditions, the switch function becomes periodic so that an extra switch necessarily occurs at the \enquote{boundary} of the sample, sharply from $1$ to $0$ (Fig.\,\ref{fig:switch_functions}(b)). Consequently $[\Lambda_2, \UE(T)]$ becomes significant also near this switch, with opposite contribution so that the index vanishes again. 
 
If the two contributions of $[\Lambda_2, \UE(T)]$ are well separated, one can also introduce a cut-off to compute the trace around one of them only, and expect to get an estimated index that is non-vanishing and coincides with $\IE$ in the thermodynamic limit.

\subsection{Effective Hamiltonian and truncation to the edge \label{sec:trunc_num}} 

Before to compute the invariant using the different cut-offs discussed above, we still need to compute the effective Hamiltonian and truncate it to get its edge version. Indeed the correct expression of $\IE(\varepsilon)$ is \eqref{defIEeps} and not \eqref{defIE}, although the previous discussion on switch functions works for both. Effective Hamiltonian $\HB^\varepsilon$ is defined through the logarithm, see \eqref{defHeff}. Since $\UB(T)$ is a $L^2\times L^2$ square matrix, the effective Hamiltonian is computed through the logarithm of a matrix. This is the only non-trivial part of the numerical implementation. All the remaining is basic linear algebra such as product of matrices. Moreover the computation of a matrix logarithm, even for large matrices, has been well studied and several efficient algorithms exist to compute it \cite{AlMohyHigham12,Loring14}.

The edge effective Hamiltonian is obtained by truncation of the bulk one, namely $\HE^\varepsilon = \widehat{\HB^\varepsilon}$. At the numerical level this simply means that we remove the off-diagonal terms of the matrix $\HB^\varepsilon$. Indeed since $\HB^\varepsilon$ is local as in \eqref{local} it has non vanishing terms only near the diagonal that decay exponentially away from it. On top of that it has far away off-diagonal terms that correspond to periodic boundary conditions. If the two contribution are not overlapping (namely if the size of the system is larger than the range of the operator), one can set the off-diagonal part to zero, leading to the same Hamiltonian but with Dirichlet boundary condition, that is $\HE^\varepsilon$. This intuitive picture is correct in one dimension but has to be carefully adapted to our two-dimensional problem since we only want Dirichlet boundary condition in direction 1 whereas direction 2 remains periodic.

We can then check that the time evolution $\ee^{-\ii T \HE^\varepsilon}$ has the same spectrum than the one in Fig.\,\ref{fig:num_spectra}(b) but without any edge mode in both gaps, in agreement with the discussion in Sect.\,\ref{sec:eff_vac} for the anomalous phase. In this example the effective vacua is present to remove contributions from the delocalized bands only.

\subsection{Numerical estimate of the index \label{sec:IEnum}}

\begin{figure}[htb]
	\centering
	\includegraphics{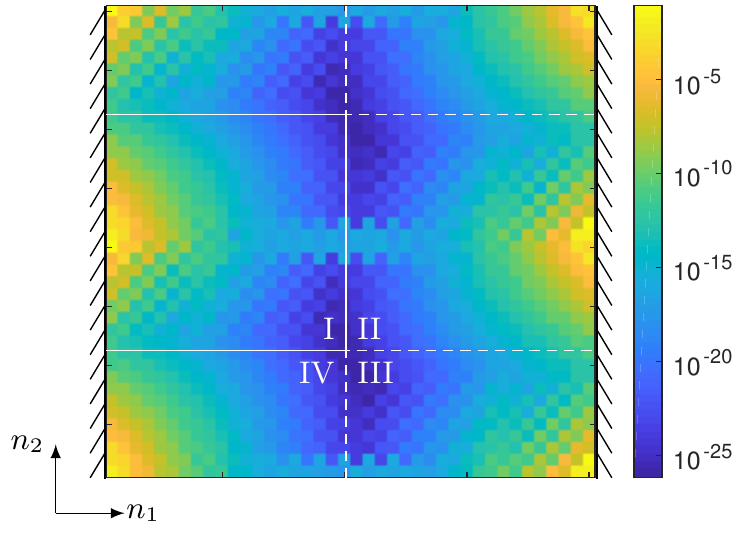}
	\caption{Diagonal kernel $|\Delta|_{\vn,\vn}$ where $\vn=(n_1,n_2) \in [1,L]^2$ for $J=1.875\pi/T$, $\delta=1.6\pi/T$ and $L=40$. Direction $1$ has Dirichlet boundary condition with two edges at $n_1=1$ and $n_1=L$, and direction $2$ has periodic boundary condition with $n_2=1$ and $n_2=L$ identified. The colorbar is the logarithmic amplitude, so that the kernel is significant only near the edges and the switches, dividing the sample into four equal areas denoted from I to IV. Each one leads to a numerical index according to \eqref{defIEnum}. \label{fig:diagonalkernel}}
\end{figure}

We have now everything to propose a numerical version of the edge index. In what follows we always assume for simplicity that the branch cut is taken at $\varepsilon=\pi$ and we look at the invariant $\IE(\pi)$ in the corresponding gap. Consider the operator
\begin{equation}
\Delta := \UE^*(T)[\Lambda_2, \UE(T)]- \ee^{\ii T \HE^\pi}[\Lambda_2,\ee^{-\ii T \HE^\pi}]
\end{equation}
that equals the one appearing in \eqref{defIEeps} by expanding the commutators, so that by definition $\IE(\pi) =\Tr(\Delta)$. The trace being the sum of diagonal elements, we focus only on $\Delta_{\vn,\vn}$ and consider its numerical version where $\vn \in [1,L]^2$. A typical amplitude of $|\Delta_{\vn,\vn}|$ is illustrated in Fig.\,\ref{fig:diagonalkernel}. Note that the amplitude scale is logarithmic so that $\Delta_{\vn,\vn}$ is significant only in a few regions.

This comes from the following facts, discussed above
\begin{itemize}
	\item Because of operator $[\Lambda_2, \cdot]$ then $\Delta_{\vn,\vn}$ is confined near the switches of $\Lambda_2$, namely around $n_2=L/2$ and $n_2=L+1=1$. 
	\item Because we consider the relative evolution with respect to the effective one, that coincide in the bulk, $\Delta_{\vn,\vn}$ is confined near the two edges of the sample. 
\end{itemize}
In each case, the two contributions compensate so one has to pick one of them only. Thus we divide the sample into four equal areas denoted from I to IV in Fig.\,\ref{fig:diagonalkernel}, according to the choice of switch and edge. For each area we define the cut-off $Q_A$ for $A \in \{\mathrm I,\ldots,\mathrm IV\}$, a diagonal operator that is 1 in area $i$ and 0 outside. The numerical invariant is then defined as
\begin{equation}\label{defIEnum}
\widetilde \IE(\pi,Q_A) := \Tr(\Delta Q_A) = \sum_{\vn \in A} \Delta_{\vn,\vn}
\end{equation}
In the thermodynamic limit when $L\rightarrow \infty$ region $\mathrm I$ becomes similar to the half-infinite space described in Sect.\,\ref{sec:quantized_pumping} so that $\widetilde \IE(\pi,Q_A)$ coincides with $\IE(\pi)$. The invariant computed in the other regions should also coincide with $\IE$ up to a sign, since the edge orientation or the switch has been reversed there.

The computation of the index is actually pretty accurate even for small sizes, as illustrated in Table \ref{tab:invariant} for the case where $J=1.875 \pi/T$ and $\delta=1.6 \pi/T$. The theoretical value of the invariant in that case $\IE(\pi)=-1$, that we get numerically at a precision scaling with the size of the system

\begin{table}[htb]
\begin{tabular}{|c||c|c|c|c|}
	\hline
	$L$ & $8$ & $16$ & $32$ & $48$ \\
	\hline
	$|\widetilde \IE(\pi,Q_\mathrm{I}) - (-1)| \leq $&  $1.10^{-2}$ & $1.10^{-4}$ & $1.10^{-8}$ & $1.10^{-12}$ \\
	\hline
	$\Delta \IE(\pi,Q_I) \leq$ & $8.10^{-4}$ & $1.10^{-5}$ & $2.10^{-9} $&  $2.10^{-13}$ \\
	\hline
\end{tabular}
\caption{Numerical index distance to its theoretical value. The precision scales with the size of the system. The last line is the standard deviation of the invariant for a large number of disordered configurations.\label{tab:invariant}}
\end{table}

Note that this invariant is computed for a given disordered configuration and does not require any average on the disorder so that the computation can be done in a few seconds (resp. minutes) for $L=8$ (resp. for $L=48$) on a simple computer. In the last line of Table \ref{tab:invariant} we give the standard deviation of the index when computed for a large number of disordered configurations. As expected, one can also check numerically that $\IE(\pi,Q_\mathrm{I}) \simeq - \IE(\pi,Q_\mathrm{II}) \simeq \IE(\pi,Q_\mathrm{III}) \simeq - \IE(\pi,Q_\mathrm{IV})$ within the same order of magnitude.

\begin{figure}[htb]
	\centering
	\includegraphics{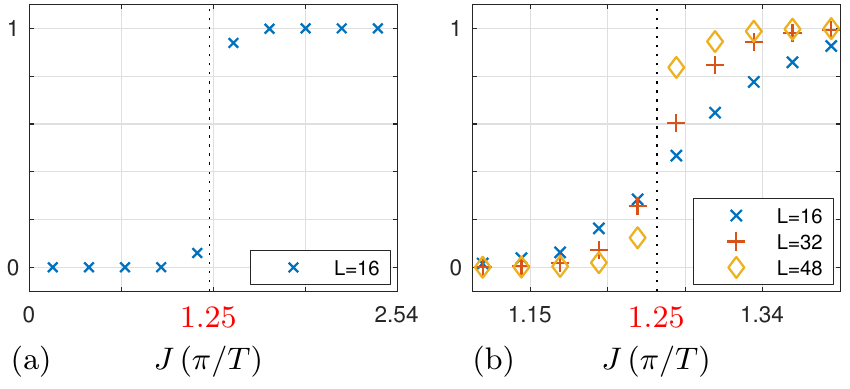}
	\caption{Numerical invariant $|\widetilde \IE(\pi,Q_\mathrm{I})|$ with respect to $J$ (in $\pi/T$ units) for $\delta = 1.6 \pi/T$. (a) Away from the transition $J^*=1.25\pi/T$ we get 0 or 1 with good precision. (b) In a narrow region near $J^*$, the index converges to the step function in the thermodynamic limit $L \rightarrow \infty$. The fluctuations due to disorder are of the size of the markers.  \label{fig:transition}}
\end{figure}

Then we compute the invariant for several values of $J$ and look at the topological transition. Away from $J^* = 1.25 \pi/T$ where the gap closes, the estimation is pretty accurate for a large range of $J$ (Fig.\,\ref{fig:transition}(a)). The numerical index moves away from integer values when looking close to the transition, but becomes more accurate and converges to the step function by increasing the size $L$ of the sample (Fig.\,\ref{fig:transition}(b)).

Notice that even near the transition, the fluctuations due to disorder remain small. The fact that $\widetilde \IE$ becomes less accurate here is actually because the range $1/\mu$ of the local operators (see \eqref{local}) goes to infinity when the gap of $\UB(T)$ closes. Thus the hypotheses that there is no overlap between the two edges or the two switches, discussed in Sect.\,\ref{sec:switch_per} or between the diagonal and far off-diagonal terms related to the truncation procedure from Sect.\,\ref{sec:trunc_num}, become less and less valid when the size of the gap decreases. Increasing the size of the system decreases the different overlaps and thus restores the hypotheses (Fig.\,\ref{fig:transition}(b)).

Finally this cut-off procedure works similarly to define $\widetilde \II$, a numerical version of the interface index, and one can check numerically that $\IE(\varepsilon) = \II(\varepsilon)$. However the latter index was introduced to interpret $\HE^\varepsilon$ as an effective vacuum rather than for computational purposes. Furthermore we claim that expression \eqref{defIEeps} of $\IE(\varepsilon)$ is slightly simpler and more efficient to be implemented numerically, but since it might also be of independent interest we describe the numerical interface index in App.~\ref{app:num_int}. Notice that $\widetilde \IE$ and $\widetilde \II$ being computations in real space with opposite contributions that compensate, this approach bares similarity with the bulk invariant defined in \cite{MazzaroResta17}, although it was for translation-invariant static systems. 

\subsection{Application to a non-anomalous case}

\begin{figure}[htb]
	\includegraphics{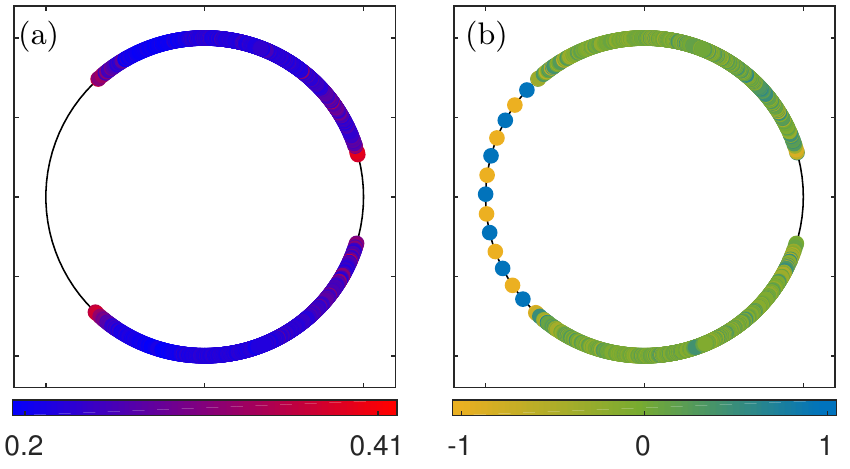}
	\caption{\label{fig:non_anomalous} Numerical spectra of a disordered configuration for $J=1.5\pi/T$, $\delta=0.5\pi/T$ and $L=40$ in the case where the random potential is present all the time. (a) Bulk spectrum with two gaps.  (b) Edge spectrum: one gap is filled with edge states whereas the other is not. The system is not anomalous.}
\end{figure}

As mentioned before, the switch-function formalism works for any Floquet system for which $\UB(T)$ has a spectral gap. Following \cite{RudnerPRX13} (that was without disorder), we modify the model of Sect.\,\ref{sec:model} by implementing the on-site random potential \eqref{defH5} at all time. This means we replace $H_i$ by $H_i + H_5$ for $i=1,\ldots,4$ and leave $H_5$ unchanged. The corresponding bulk and edge spectra are given in Fig.\,\ref{fig:non_anomalous}.

Similarly to the previous case the bulk spectrum has two gaps, but in the edge picture only one is filled with edge modes. This is a numerical realization of Fig.\,\ref{fig:bulk_edge_spectra}. In particular the Chern numbers of the bulk bands are non-vanishing, and the system is not anomalous. One can check (not shown) that the spectrum of the effective vaccum is in agreement with Fig.\,\ref{fig:effective_spectrum}. Moreover, because of non-zero Chern numbers, the localization length of some bulk states is infinite in the thermodynamic limit (if the disorder is not too strong). We can anyway compute the edge index that is already quantized for small sizes, as we can see in Table \ref{tab:invariant_NA}.

\begin{table}[htb]
	\begin{tabular}{|c||c|c|c|c|}
		\hline
		$L$ & $8$ & $16$ & $32$ & $48$ \\
		\hline
		$|\widetilde \IE(\pi,Q_\mathrm{I}) - (-1)| \leq $&  $5.10^{-2}$ & $2.10^{-3}$ & $5.10^{-6}$ & $2.10^{-8}$ \\
		\hline
		$\Delta \IE(\pi,Q_I) \leq$ & $9.10^{-3}$ & $6.10^{-4}$ & $2.10^{-6} $&  $1.10^{-8}$ \\
		\hline
	\end{tabular}
	\caption{Numerical index distance to its theoretical value and standard deviation for a large number of disordered configurations for the non-anomalous model.\label{tab:invariant_NA}}
\end{table}

Note that the Chern numbers do not entirely capture the topology of a non-anomalous Floquet model. Moreover their physical interpretation is less clear than in static topological insulators. Thus the computation and physical interpretation of the switch-function edge index appears also helpful in that case in order to characterize the topology of disordered models. The transition from non-anomalous to anomalous phase could also be studied within this formalism.

\section{Discussion \label{sec:ccl}}

In this paper we focused on the edge properties of Floquet topological phases. We showed that the quantized pumping within a Floquet cycle, first noticed for the anomalous Floquet-Anderson insulator phase, is actually a general phenomena for any gapped bulk one-period evolution. This pumping is somewhat hidden among other transport processes of the dynamics, that have to be subtracted in some way in order to observe the topological part at the edge of the sample. This regularization is provided by expression \eqref{defIEeps} of $\IE$ coming from the relative dynamics with respect to an effective one. This expression, based on the switch function formalism, is deterministic and works for every disordered configuration of the sample, as long as the bulk evolution after one period has a spectral gap. 

The effective dynamics at the edge, given by $\HE^\varepsilon$, can be easily interpreted in the interface picture that is dual to the previous one. There $\HE^\varepsilon$ appears as an effective vacuum that, when placed next to the original sample, reproduces the quantized pumping at the interface, relatively to the transport processes that could occur in the bulk. This effective vacuum depends on the choice of spectral gap  and actually appears as a way to select a given gap among several of the bulk evolution, in analogy with chemical potential that shifts Fermi energy in static topological insulators. However there is no notion of ground state here and the effective vacuum is actually independent of the quasi-energy inside a given gap, so the analogy should be taken with care. Nevertheless when Floquet operator has several gaps with different invariants effective vacua are definitely a way to select and observe the corresponding edge modes in each of them.

These effective vacua are defined as the truncation of the logarithm of the bulk evolution, and thus might be cumbersome to handle in practice. However most of their properties are immediate and simple: they are time-independent and local. In particular they can always be well approximated by finite range static Hamiltonian. Moreover the spectrum of an effective vacuum $\HE^\varepsilon$ is easily deductible from the bulk evolution. By construction it has no edge mode in the gap around $\varepsilon$ and has the same bands than the edge evolution but with different edge modes according to the Chern numbers of the bulk evolution. In the particular case of anomalous phases it has no edge mode at all.

We finally implemented these concepts in a numerical framework on a generic model, leading to an accurate estimation of the edge index $\IE$, without average and for any disordered configuration. Once fixed the issue with switch functions in a periodic setting, the complexity of the algorithm is reduced to compute the logarithm of a matrix that is quite well implemented nowadays. The rest of the code is basic linear algebra and computes the edge invariant for any disordered system from the input of $\UE(T)$ and $\UB(T)$. The algorithm becomes less relevant near the topological transition, where the range of the system goes to infinity and thus breaks the hypotheses on which the code is built. Away from the transition it is quite accurate and efficient.

This works open several interesting perspectives. The effective vacua and the interface with them provide a mechanism to select and observe the topological properties associated to one of the spectral gaps of a Floquet topological insulator. Moreover we also learned from the definition of the index that the quantized pumping occurring at the edge (or equivalently at the interface) can only be measured up to a regularization that removes the eventual contributions from the bands and other gaps. In other words the word \enquote{insulator} seems slightly inappropriate here. In contrast with the static case, the spectral bands of a Floquet evolution also contribute within a cycle and may lead to transport processes, topological or not. Somehow the dynamics of periodically driven system seems to be richer since the transport of electrons involves several distinct contributions. The effective vacua appears as a way to disentangle them. 

Furthermore the index $\II$ of an interface given in \eqref{defII} is also  of independent interest, since it works for any two edge evolutions that coincide in the bulk and is independent on the way they are glued together at the interface. Here it was mostly used to give an interpretation of $\HE^\varepsilon$ but in principle one could take two bulk evolutions such that $\UBo(T) = \UBt(T)$ and expect a topological pumping at the interface. A system with a continuous parameter that drives the topological transition might have two distinct parameter values with a coinciding bulk evolution, see for example \cite{DelplaceFruchartTauber17} in the context of oriented scattering networks. In that case the second evolution plays the role of a \emph{dynamical vacuum} placed next to the first one, in contrast with an effective vacuum that is time independent. A dynamical vacuum might be however easier to implement in practice.

Finally, the numerical implementation of the index $\IE$, that has an illustrative purpose here, is also of independent interest. It paves the way for a general procedure to estimate any index in the context of disordered topological insulators. Coming from functional analysis, the formalism of switch function has been underestimated in the physics literature whereas it does not require any strong knowledge of the underlying mathematical theory. Roughly speaking for a switch function $\Lambda_i$ in direction $i$, $[\Lambda_i, \cdot]$ replaces quasi momentum derivative $\partial_{k_i}$ of translation invariant system. In our case, comparing \eqref{defIE} and \eqref{IEwinding} we have explicitly
\begin{equation}
\dfrac{1}{2\pi \ii} \int \dd k_i \Tr_{j} ( * \, \partial_{k_i} * ) \quad  \leftrightarrow \quad  \Tr_{i,j} (* \, [\Lambda_i, *])
\end{equation}
but this can be generalized to other dimensions or extra symmetries, at least as a formal expression. It is then possible to rigorously define the index in terms  of switch functions without requiring translation-invariance. This has been done for the Floquet bulk index \cite{GrafTauber17}, but also originally for the Chern number \cite{AvronSeilerSimon94} and other static topological insulators \cite{GrafShapiro18,MarcelliPanatiTauber18}. In any case the cut-off procedure of this paper provides a simple algorithm to immediately estimate these quantities even if they are formally defined only. Thus in principle it is possible to generalize and study the robustness to disorder of any quantity initially defined in terms of quasi-momentum derivatives.  The issue of strong disorder and mobility gap might be also studied within this formalism.

\appendix

\section{Derivation of edge index expression \label{app:IEeps}}

The relative edge Hamiltonian given by \eqref{defHErel} generates the following time evolution
\begin{equation}\label{defUErel}
\UE^\mathrm{rel}(t) = \left\lbrace\begin{array}{lll}
\UE(2t) & \mathrm{for} & 0 \leq t \leq T/2\\
\ee^{- \ii (T-2t) \HE^\varepsilon} \UE(T) & \mathrm{for} & T/2 \leq t \leq T
\end{array}\right. 
\end{equation}
In particular $\UE^\mathrm{rel}(T) = \ee^{ \ii T \HE^\varepsilon} \UE(T)$. Since $\UB^\mathrm{rel}(T)=1$ we can use expression \eqref{defIE} with $\UE^\mathrm{rel}(T)$ to define
\begin{equation}\label{defIEepsapp}
\IE(\varepsilon) := \Tr\Big( \UE^*(T)\ee^{ -\ii T \HE^\varepsilon} \Lambda_2 \ee^{ \ii T \HE^\varepsilon}\UE(T) - \Lambda_2 \Big)
\end{equation}
We then use the invariance under continuous deformation of the index and consider the homotopy $V(s) = \ee^{-\ii s T \HE^\varepsilon} \UE^\mathrm{rel}(T) \ee^{\ii s T \HE^\varepsilon}$ for $s \in [0,1]$  so that $V(0)=\UE^\mathrm{rel}(T)$ and $V(1) = \UE(T)\ee^{ \ii T \HE^\varepsilon}$. Importantly, this homotopy preserves $\UE^\mathrm{rel}(T) = 1 + D(T)$, namely $V(s) = 1 + D(T,s)$ with $D(T,s)$ confined near the edge. Thus the edge index is well defined and  remains constant for every $s$, so that at $s=1$
\begin{equation}
\IE(\varepsilon) = \Tr\Big( \ee^{ -\ii T \HE^\varepsilon}\UE^*(T) \Lambda_2\UE(T) \ee^{ \ii T \HE^\varepsilon} - \Lambda_2 \Big)
\end{equation}
By conjugating the entire expression under the trace by $\ee^{ \ii T \HE^\varepsilon}\, \cdot \,\ee^{-\ii T \HE^\varepsilon}$ we get expression \eqref{defIEeps} for $\IE(\varepsilon)$. Note that definition \eqref{defIEepsapp} is perfectly valid for $\IE(\varepsilon)$ but the pumping interpretation is less obvious as the operators appear in the wrong order.

\section{Edge-interface correspondence \label{app:IE=II}}

Consider the sharp interface given by \eqref{sharpH}. We denote this Hamiltonian by $H_\# = \HE^\varepsilon \# \HE$, meaning that $\HE^\varepsilon$ is on the left and $\HE$ on the right of the sample. In that case $H_\#$ is composed of two disconnected blocks, then so is the corresponding evolution. In particular $U_\#(T) = \ee^{-\ii T \HE^\varepsilon} \# \UE(T)$. For a general interface $\HI$ we claim that the gluing condition $H_{\mathrm{int}}$ does not change much since $\HI$ is continuously deformable to $H_\#$, so that in particular $\UI(T) \simeq U_\#(T)$.

Then notice that $\UB(T) = \ee^{-\ii T \HB^\varepsilon}$ by construction, so we can use the later expression in the interface index definition \eqref{defII}. Similarly we can consider $\HB^\varepsilon$ as a trivial gluing of $\HE^\varepsilon$ on both half of the space, namely $\HB^\varepsilon \simeq \HE^\varepsilon \# \HE^\varepsilon$. Of course the equality is approximatively true only as the r.h.s corresponds to a disconnected interface, but we claim that the error is small when $\HB^\varepsilon$ is local. Furthermore $\HB^\varepsilon$ can be continuously deformed to $\HE^\varepsilon \# \HE^\varepsilon$. As a consequence one has $\ee^{-\ii T \HB^\varepsilon} \simeq \ee^{-\ii T \HE^\varepsilon} \# \ee^{-\ii T \HE^\varepsilon}$. Consequently,
\begin{equation}
\UB^*(T) \UI(T) \simeq 1 \, \# \, \ee^{\ii T \HE^\varepsilon} \UE(T)
\end{equation}
Then we rewrite the interface index \eqref{defII}, by cyclicity of the total trace,
\begin{equation}
\II(\varepsilon) = \Tr \Big( \UI^*(T) \UB(T)\Lambda_2 \UB^*(T) \UI(T) - \Lambda_2 \Big)
\end{equation}
and realize that the left-half contribution (namely $1$) cancels out from this formula. Thus we are left with the trace on the right-half space, that is the edge space, and the expression is exactly the edge index definition \eqref{defIEepsapp}. The approximations made in this sharp interface computation can be removed through continuous deformations and leave the indices unchanged \cite{GrafTauber17}, leading to \eqref{IE=II}.

\section{Numerical interface index \label{app:num_int}} 

\begin{figure}[htb]
	\centering
	\includegraphics{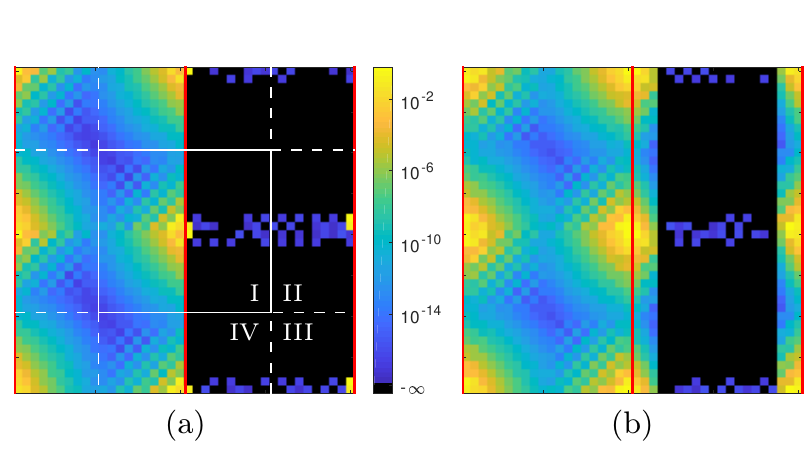}	
	\caption{\label{fig:interface_num} Amplitude (in logarithmic scale) of $|\Delta_{\vn,\vn}|$ for interfaces between $\HE^{\varepsilon}$ on the left and $\HE$ on the right: disconnected interfaces (a), and nearest neighbor hopping at the interfaces (b). In each case the amplitude is significant near the interfaces $n_1=L/2$ and $n_1=1=L$ (red lines) and near the switch in direction $n_2$, namely $n_2=L/2$ and $n_2=1=L$. This divides the sample into four equivalent areas on each of which the numerical interface invariant can be computed. The parameters are the same as in Fig.\,\ref{fig:diagonalkernel}.}
	
\end{figure}

Similarly, it is also possible to implement a numerical estimate of the interface index. In that case the edge Hamiltonian is given by the lower-right block of $\HB$, which restricts the bulk sample to the right half-space with Dirichlet boundary condition in direction 1 at $n_1=L/2$ and $n_1=L$. Then we take the upper-left block of the effective bulk Hamiltonian $\HB^\varepsilon$ (computed as the logarithm of $\UB(T)$), that restricts it to the left half-space with Dirichlet boundary condition at $n_1=1$ and $n_1=L/2$. When put together, these two pieces constitute a sharp (or disconnected) interface between $\HE^\varepsilon$ and $\HE(t)$ at $n_1=L/2$ and $n_1=1=L+1$ by periodicity. The corresponding Hamiltonian $\HI(t)$ is block diagonal and generates evolution $\UI(t)$. The operator appearing in the definition of the interface index is $\Delta=\UI^*(T)[\Lambda_2,\UI(T)] - \UB^*(T)[\Lambda_2,\UB(T)]$ where $\UB$ is the bulk evolution on the whole sample with periodic boundary conditions in both directions. The diagonal elements $|\Delta_{\vn,\vn}|$ are represented in Fig.\,\ref{fig:interface_num}(a) for a sharp interface.

The sample splits again into four equivalent regions because $\Delta$ is confined near the $\Lambda_2$-switches and near the two interfaces, each of one giving a numerical index by applying the cut-off procedure of Sect.\,\ref{sec:IEnum}: $\widetilde \II(\varepsilon, Q_A)$ is defined as in \eqref{defIEnum} but with the areas of Fig.\,\ref{fig:interface_num}(a) instead. We recover the same accuracy than on Table \ref{tab:invariant} for the edge index. Note that to the right of the interface almost all the elements of $|\Delta_{\vn,\vn}|$ are zero. This is because the model described in Sect.\,\ref{sec:model} is actually a sequence of isolated two-level systems, so that away from the interface the evolution $\UE$ is not only local but finite range. On the other side the effective Hamiltonian $\HE^\varepsilon$ is defined in terms of a logarithm, which is not finite range even when $\UB(T)$ is. This is why on the left of the interface operator $\Delta$ has a richer structure, even though these matrix elements are very small away from the significant regions.

Finally we can connect the two parts in a more physical way. We add for example a time-independent nearest neighbor hopping term of amplitude $J$ at the interfaces $n_1=L/2$ and $n_1=1=L+1$, allowing for exchanges between the two halves. The corresponding operator $\Delta$ is represented in Fig.\,\ref{fig:interface_num}(b), where we observe that the two dynamics have been mixed together along the interface within a cycle. The estimated index anyway leads to the same integer value, but with a slightly lower precision than in the previous case. The orders of magnitude of Table \ref{tab:invariant} are strictly recovered when we take a rectangle sample of size $2 L\times L$ instead.

\end{document}